# Sustainable Cloud Computing: Foundations and Future Directions

Rajkumar Buyya and Sukhpal Singh Gill

Cloud Computing and Distributed Systems (CLOUDS) Laboratory
School of Computing and Information Systems
The University of Melbourne, Australia

rbuyya@unimelb.edu.au, sukhpal.gill@unimelb.edu.au

## Abstract

Major cloud providers such as Microsoft, Google, Facebook and Amazon rely heavily on datacenters to support the ever-increasing demand for their computational and application services. However, the financial and carbon footprint related costs of running such large infrastructure negatively impacts the sustainability of cloud services. Most of existing efforts primarily focus on minimizing the energy consumption of servers. In this paper, we devise a conceptual model and practical design guidelines for holistic management of all resources (including servers, networks, storage, cooling systems) to improve the energy efficiency and reduce carbon footprints in Cloud Data Centers (CDCs). Furthermore, we discuss the intertwined relationship between energy and reliability for sustainable cloud computing, where we highlight the associated research issues. Finally, we propose a set of future research directions in the field and setup grounds for further practical developments.

## 1. CDCs and the Challenge of Sustainable Energy

The cloud computing paradigm offers on-demand, subscription-oriented services over the Internet to host applications and process user workloads. To ensure the availability and reliability of the services, the components of Cloud Data Centers (CDCs) such as network devices, storage devices and servers should be run 24/7. Large amounts of data are created by digital activities such as data streaming, file sharing, searching and social networking websites, e-commerce, sensor networks that data can be stored as well as processed efficiently using cloud datacenters [1]. However, creating, processing, and storing each bit of data adds to the energy cost, increases carbon footprints, and further impacts the environment [2]. Due to the large consumption of electricity by CDCs, the community is facing the challenge of sustainable energy economy. The amount of energy consumed by the CDCs is increasing regularly as shown in Figure 1 and it is expected to be 8000 Tera Watt hours (TWh) by 2030 [3].

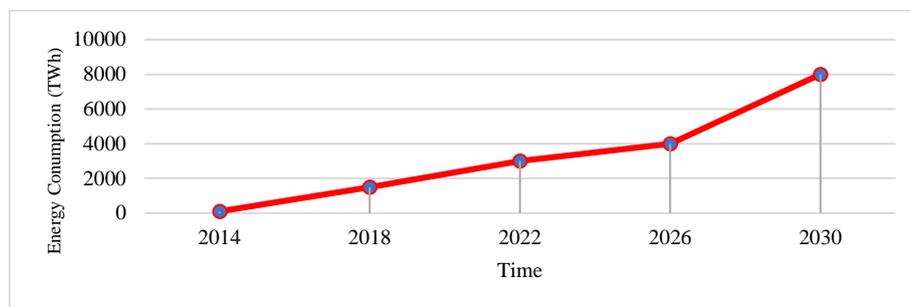

Figure 1: Energy Consumption in Cloud Datacenters

Moreover, existing energy-aware techniques mainly focus on reducing the energy consumption of the servers [4]. The other components (networks, storage, memory and cooling systems) of CDCs are consuming huge amount of energy. To improve energy efficiency of CDC, there is a need for energy-aware resource management technique for management of all the resources (including servers, storage, memory, networks, and cooling systems) in a

*Cite as:* Rajkumar Buyya and Sukhpal Singh Gill. "Sustainable Cloud Computing: Foundations and Future Directions." Business Technology & Digital Transformation Strategies, Cutter Consortium, Vol. 21, no. 6, Pages 1-9, 2018.

holistic manner. Due to the underloading/overloading of infrastructure resources, the energy consumption in CDCs is not efficient; in fact, most of the energy is consumed while some resources (i.e., networks, storage, memory, processor) sit in an idle state, increasing the overall cost of cloud services [1] [2].

In the current scenario, CDC service providers are finding other alternative ways to reduce carbon footprints of their infrastructure [4]. The prominent cloud providers such as Google, Amazon, Microsoft and IBM are hoping to power their datacenters using renewable energy sources [5]. Future CDCs are required to provide cloud services with minimum emissions of carbon footprints and heat release in the form of greenhouse gas emissions. So what are some of the concerning issues?

Well, an efficient cooling mechanism is required to maintain the temperature of datacenters, but it increases costs. Further, cooling expenses can be decreased by developing waste heat utilization and free cooling mechanisms. Location-aware ideal climatic conditions are needed for an efficient implementation of free cooling and renewable energy production techniques. Moreover, waste heat recovery locations are required to be identified for an efficient implantation of waste heat recovery prospects [6]. To enable sustainable cloud computing, datacenters can be relocated based on: i) opportunities for waste heat recovery, ii) accessibility of green resource and iii) proximity of free cooling resources. To resolve these issues and substantially reduce energy consumption of CDCs, there is a need for cloud computing architectures that can provide sustainable cloud services through holistic management of resources.

## 2. A Conceptual Model

Figure 2 shows a conceptual model for sustainable cloud computing in the form of layered architecture, which offers holistic management of cloud computing resources, to make cloud services more energy-efficient and sustainable. The four main components of proposed architecture are discussed below:

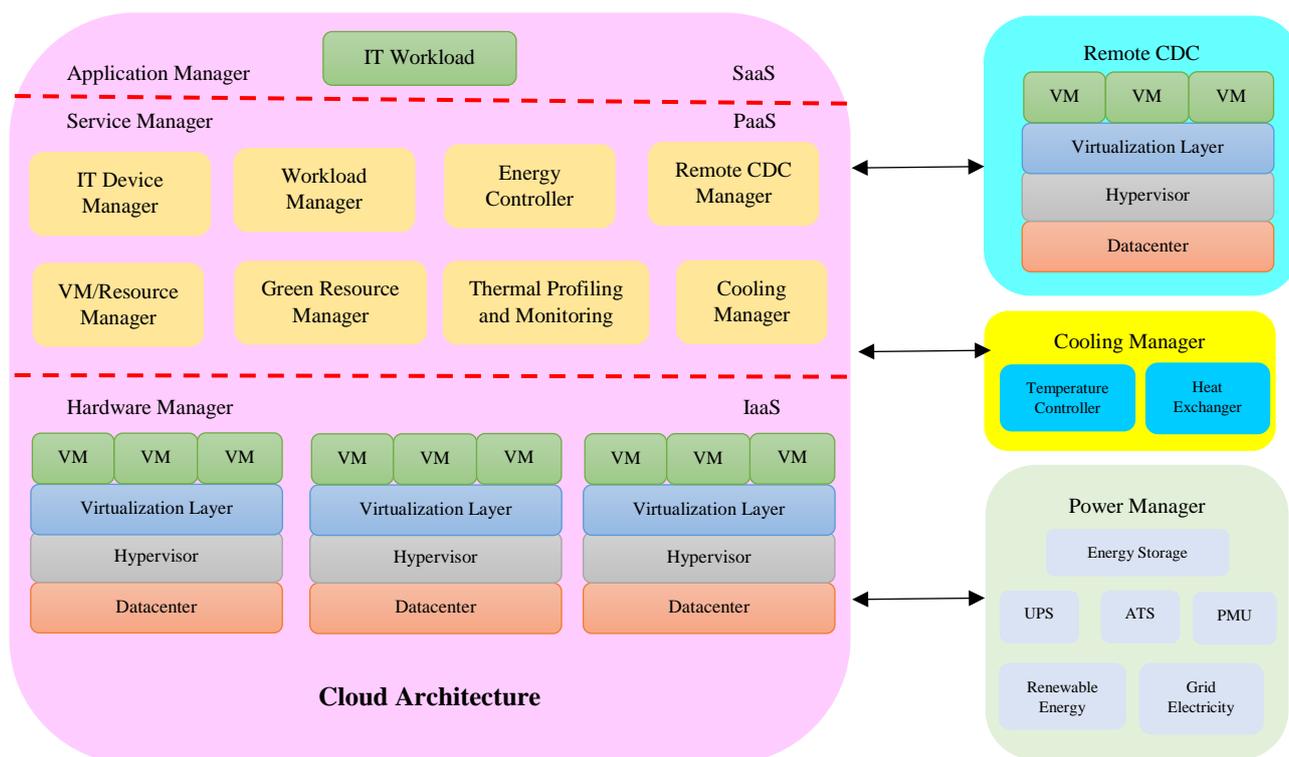

Figure 2: A Conceptual Model for Sustainable Cloud Computing

1. **Cloud Architecture:** This component is divided into three different sub-components: Software as a Service, Platform as a Service and Infrastructure as a Service.

*Cite as:* Rajkumar Buyya and Sukhpal Singh Gill. "Sustainable Cloud Computing: Foundations and Future Directions." Business Technology & Digital Transformation Strategies, Cutter Consortium, Vol. 21, no. 6, Pages 1-9, 2018.

   a) *Software as a Service (SaaS):* At this layer, application manager is deployed to handle the incoming user workloads, which can be interactive or batch style and transfer to workload manager for resource provisioning.

   b) *Platform as a Service (PaaS)*: At this layer, controller or middleware is deployed to control the important aspects of the system. *IT device manager* manages all the devices attached to cloud datacenter. *Workload manager* manages the incoming workloads from the application manager and identifies the Quality of Service (QoS) requirement for every workload for their successful execution and transfer the QoS information of workload to the VM/resource manager. *Energy controller* manages the energy consumption of cloud datacenters to ensure sustainability of cloud services. *Remote CDC manager* manages the VM migration and workload migration between local and remote cloud datacenters for effective utilization of energy. *VM/resource manager* provisions and schedules the cloud resources for workload execution based on QoS requirements of workload using physical machines or virtual machines. *Green resource manager* manages the electricity coming from power manager and it prefers renewable energy as compared to grid electricity to enable sustainable cloud environment. *Thermal profiling and monitoring* technique is used to analyze the temperature variations of the cloud datacenter based on the temperature value as monitored by thermal sensors. Cooling manager controls the temperature of the cloud datacenter at infrastructure level.

   c) *Infrastructure as a Service (IaaS):* This layer contains the information about cloud datacenters and Virtual Machines (VM). VM migrations are performed to balance the load at virtualization layer for efficient execution of workloads. The proactive temperature-aware scheduler is used to monitor the temperature variation of different VMs running at different cores. Power Management Unit (PMU) is integrated to power all the hardware executing the VMs. Dynamic Random-Access Memory stores the current states of VMs. Thermal sensor is used to monitor the value of temperature to generate an alert if temperature is higher than its threshold value and pass the message to the heat controller for further action.

2. **Cooling Manager:** Thermal alerts will be generated if temperature is higher than the threshold value and heat controller will take an action to control the temperature with minimal impact on the performance of the CDC. The electricity coming from Uninterruptible Power Supply (UPS) is used to run the cooling devices to control the temperature. District heating management is integrated in which, the temperature is controlled by using chiller plant, outside air economizer and water economizer.

3. **Power Manager:** It controls the power generated from renewable energy resources and fossil fuels (grid electricity). To enable sustainable cloud environment, renewable energy is more preferred as compared to grid energy. If there is execution of deadline oriented workloads, then grid energy can be used to maintain the reliability of cloud services. The sources of renewable energy are solar and wind. Batteries are used to store the renewable energy. Automatic Transfer Switch (ATS) is used to manage the energy coming from both sources (renewable energy and grid electricity) and forwards it to UPS. Further, Power Distribution Unit is used to transfer the electricity to all the CDCs and cooling devices.

4. **Remote CDC:** VMs and workloads can be migrated to a remote CDC to balance the load effectively.

## 3. Implication of Reliability on Sustainability

Sustainable cloud services are attracting more cloud customers and making it more profitable. Improving energy utilization, which reduces electricity bills and operational costs to enables sustainable cloud computing [13] [14]. On the other hand, to provide reliable cloud services, the business operations of different cloud providers such as Microsoft, Google, and Amazon are replicating services, which needs additional resources and increases energy consumption. To overcome this impact, a trade-off between energy consumption and reliability is required to provide cost-efficient cloud services. Figure 3 shows the research issues related to energy and reliability for sustainable cloud computing. Existing energy efficient resource management techniques consume a huge amount of energy while executing workloads, which decreases resources leased from cloud datacenters. Dynamic Voltage and Frequency Scaling (DVFS) based energy management techniques reduced energy consumption, but response time and service delay are increased due to the switching of resources between high scaling and low scaling modes. Further, reliability of the system component is also affected by excessive turning on/off servers. Power modulation decreases the reliability of server components like storage devices, memory etc. By reducing energy consumption of CDCs, we can improve the resource utilization, reliability and performance of the server.

Therefore, there is a need of new energy-aware resource management techniques to reduce power consumption without affecting the reliability of cloud services.

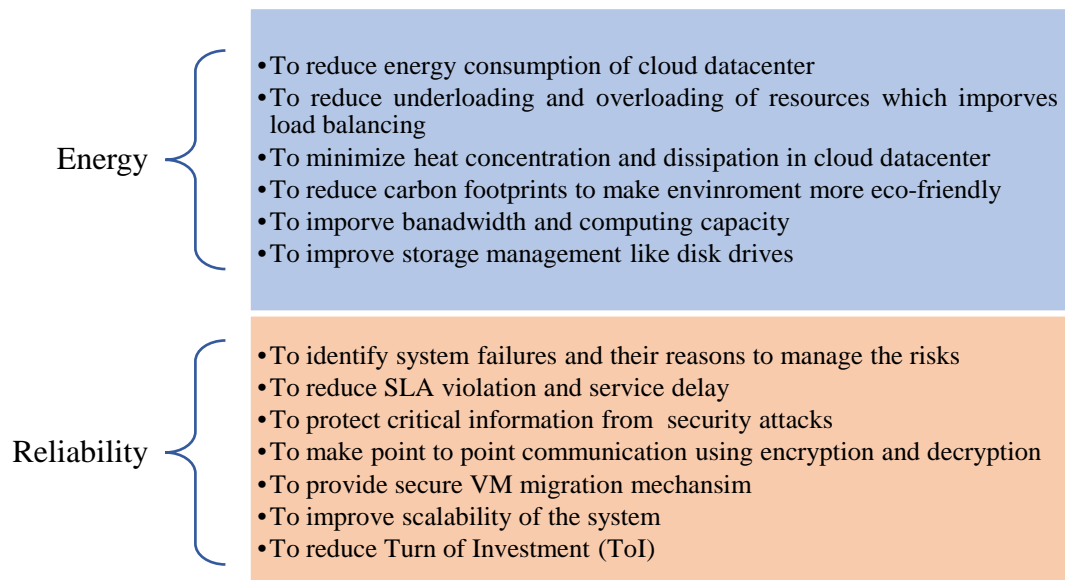

Figure 3: Research Issues related to Energy and Reliability

## 4. Areas to Investigate

The ever-increasing demand of cloud computing services that are deployed across multiple CDCs harness significant amount of power, resulting in high carbon emissions and affects the environment. In sustainable cloud computing, the cloud datacenters are powered by renewable energy resources by replacing the conventional fossil fuel based grid electricity or brown energy to effectively reduce the carbon emissions [7]. Employing energy efficiency mechanisms also makes cloud computing sustainable by reducing carbon footprints to a great extent. Waste heat utilization from heat dissipated through servers and employing mechanisms for free cooling of the servers make the CDCs sustainable [8].

Thus, sustainable cloud computing covers the following elements in making the datacenter sustainable [9]: i) using renewable energy instead of grid energy generated from fossil fuels, ii) utilizing the waste heat generated from heat dissipating servers, iii) using free cooling mechanisms and iv) using energy efficient mechanisms. All of these factors contribute in reducing the carbon footprints, the operational cost, and the energy consumption. Research issues related to sustainable cloud computing have been organized into seven categories: application model, resources targeted in energy management, thermal-aware scheduling, virtualization, capacity planning, renewable energy and waste heat utilization as shown in Figure 4. Although couple of works [3-6] [8-14] have explored issues in sustainable cloud computing, there existing many open issues in context of models for application composition, resources targeting for energy management, scheduling, capacity planning, harnessing of renewable energy and heat generated by resources.

### 4.1 Application Model
In sustainable cloud computing, the application model plays a vital role and the efficient structure of an application can improve the energy efficiency of cloud datacenters. Applications models can be data parallel, function parallel and message passing [1] [2]. Data parallel model is a form of parallelization across multiple processors in parallel computing environments, which focuses on distributing the data across different nodes, which operate on the data in parallel. The examples of data parallel model are Map-Reduce model and bag of task or parameter sweep model. Function parallel model is a form of parallelization of computer code across multiple processors in parallel computing environments, which focuses on distributing tasks concurrently, which are performed by processes or threads across different processors. The examples of data parallel model are thread and task model. Message Passing Interface provides a communication functionality between a set of processes, which are mapped to nodes or servers in a language-independent way and it encouraged development of portable and scalable large-scale parallel applications.

**4.2 Resources Targeted in Energy Management**

Many solutions have been proposed to improve energy efficiency of CDCs. The energy consumption of processor, memory, storage, network and cooling of cloud datacenters is reported as 45%, 15%, 10%, 10% and 20% respectively [1] [4] [10]. The processor is consuming huge amount of energy followed by cooling management. Power regulation approaches increase energy consumption during workload execution, which affects the resource utilization of CDCs. Further, DVFS solved the problem of resource utilization but switching of resources between high scaling and low scaling modes increases response time and service delay, which violates the SLA. Putting servers in sleeping mode or turning on/off servers affects the reliability of the system components such as storage. The objective of improving energy efficiency of cloud datacenters affects the resource utilization, reliability and performance of the server.

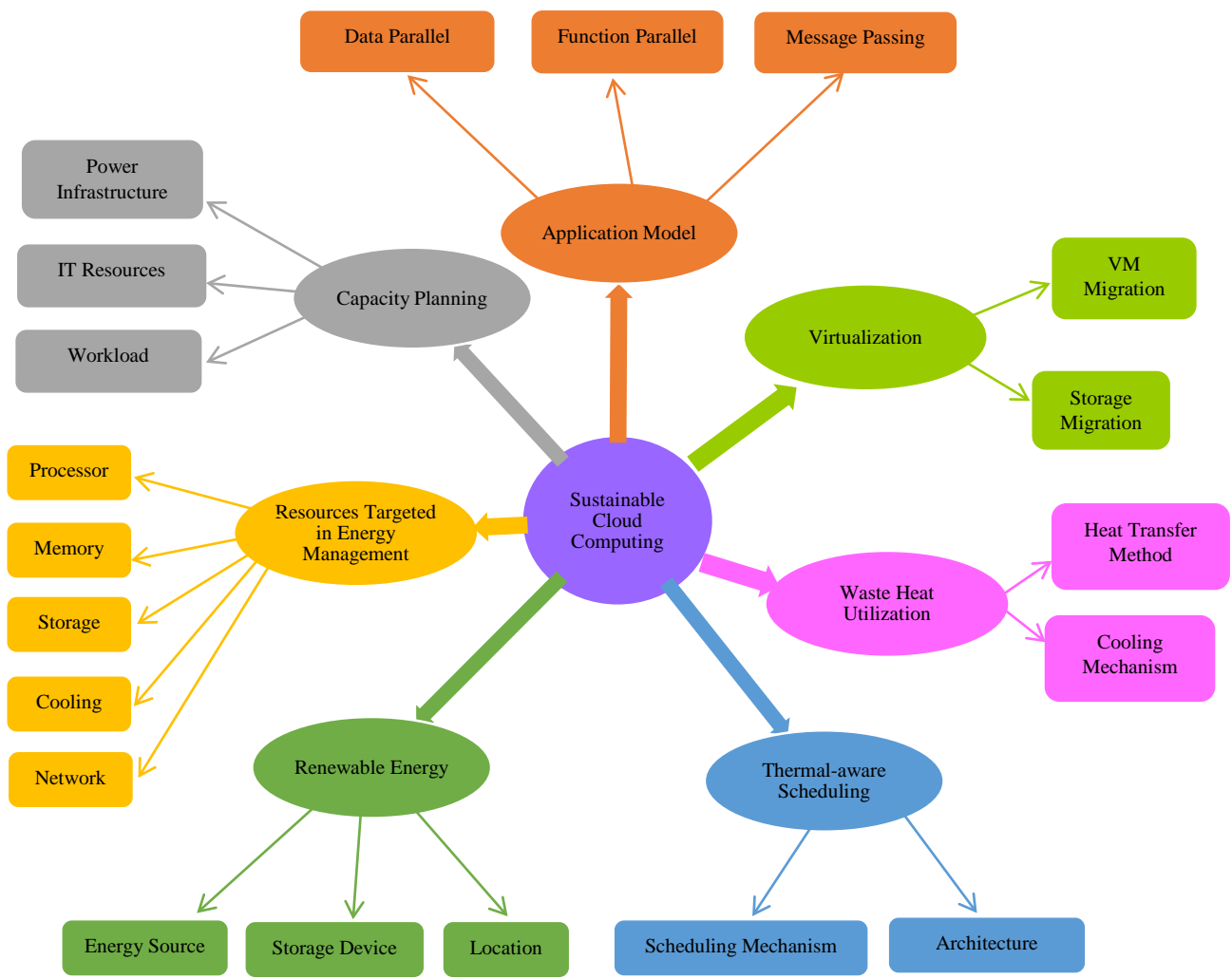

Figure 4: Research Issues in Sustainable Cloud Computing

The bin packing solution has been used in existing energy-aware resource management techniques to allocate resources for the execution of workloads. Resource allocation faces two problems: i) under-utilization of resources (resources are reserved in advance, but resource requirement is lower than resources availability, which increases cost) and ii) over-utilization of resources (a large number of workloads are waiting for execution due to unavailability of sufficient amount of resources). There is a number of methods proposed to control energy consumption by scaling down the high voltage supply, but the best way is to exploit the stall time.

**4.3 Thermal-aware Scheduling**

The important components of thermal-aware scheduling are architecture and scheduling mechanisms. Architecture can be single-core or multi-core while scheduling mechanism can be reactive or proactive. Heating problem during execution of workloads reduces the efficiency of cloud datacenters. To solve the heating problem of CDCs, thermal-aware scheduling is designed to minimize the cooling setpoint temperature, hotspots and thermal gradient [5]. Thermal-aware scheduling is economical and effective as compared to heat modelling. The energy consumption of CDCs can be minimized by activating those servers that are adjacent to each other in a rack or chassis, but power density increases, which creates heat concentration. To solve this problem, a cooling mechanism is required. There is a need of effective thermal-aware scheduling techniques, which can execute workloads with minimum heat concentration and dissipation [6]. This also reduces the load of cooling mechanism and electricity can be saved. Complexity of scheduling and monitoring is increased due to variation of temperatures of the servers in CDC, which also causes vagueness in thermal profiling. To solve this problem, there is a need of dynamically updated thermal profiles instead of static, which will be updated automatically, and provide more accurate values of temperature. Existing thermal-aware techniques focused on reducing Power Usage Efficiency (PUE) can be found, but a reduction in PUE may not reduce the Total Cost of Ownership (TCO).

### 4.4 Virtualization

During the execution of workloads, VM migration is required to balance the load effectively to utilize renewable energy resources in decentralized CDCs. Due to the lack of on-site renewable energy, VM techniques migrate the workloads to the other machines distributed geographically. VM technology also offers migration of workloads from renewable energy based cloud datacenters to the cloud datacenters utilizing the waste heat at another site. To balance the workload demand and renewable energy, VM based workload migration and consolidation techniques provide virtual resources using few physical servers. To optimize the performance of virtualization, storage from one running server to another can be migrated without affecting the workload execution of VM [7]. Waste heat utilization and renewable energy resource alternatives are harnessed by VM migration techniques to enable sustainable cloud computing. It is a great challenge for VM migration techniques to improve energy utilization and network delay while migrating workloads between distributed resources geographically. Increasing the size of VM consumes more energy, which can increase service delay. To solve this problem, point to point communication is required for VM migration using WAN [8].

### 4.5 Capacity Planning

Cloud service providers must involve an effective and organized capacity planning to attain a solid Return On Investment (ROI). The capacity planning can be done for power infrastructure, IT resources and workloads. SLA should define service quality parameters to ensure backup and recovery, storage and availability that improves user satisfaction and attracts more customers in future [9]. There is a need to consider important utilization parameters per application to maximize the utilization of resources through virtualization by finding the applications, which can be merged. Merging of applications improves resource utilization and reduces capacity cost. For efficient capacity planning, cloud workloads should be analysed before execution to finish its execution for deadline-oriented workloads. To manage power infrastructure effectively, VM migration should be provided for migration of workloads or machines to successfully complete the execution of workloads with minimum usage of resources, which improves the energy efficiency of cloud datacenters. There is a need of effective capacity planning for data storage and their processing effectively at lower cost.

### 4.6 Renewable Energy

The energy source (solar or wind), the energy storage device (net-metering or batteries) and the location (off-site or on-site) are important components of renewable energy, which can be optimized. The main challenges of renewable energy are unpredictability and capital cost of green resources. Workload migration and energy-aware load balancing techniques addressed the issue of unpredictability in supply of renewable energy [10]. Mostly, sites of commercial cloud datacenters are located away from abundant renewable energy resources. Consequently, moveable cloud datacenters are required to place nearer the renewable energy sources to make cost effective [13]. Further, Carbon Usage Efficiency (CUE) can be reduced by adding more renewable energy resources. Adoption of renewable energy in cloud datacenters has a research challenge of high capital cost.

### 4.7 Waste Heat Utilization

The cooling mechanism and heat transfer model plays an important role to utilize waste heat effectively. Due to consumption of large amounts of energy, CDCs are acting as a heat generator. The vapor-absorption based cooling systems of CDCs can use waste heat then it utilizes the heat while evaporating [11] [12]. Vapor-absorption based free cooling techniques can make the value of PUE ideal by neutralizing cooling expenses. Low temperature areas can use the heat generated by CDC for heating facilities. Power densities of servers are increasing by using stacked and multi-core server designs, which further increases the cooling costs. The energy efficiency of CDCs can be improved by reducing the energy usage in cooling. There is a need to change the location of cloud datacenters to reduce cooling costs and it can be done through placing the CDCs in areas that have availability of free cooling resources.

## 5. Summary

We identified the need for and issues in sustainability of cloud computing environments. We proposed a conceptual model for holistic management of resources to decrease carbon footprints of cloud datacenters, which makes cloud services more energy-efficient and sustainable. Holistic management improves the energy efficiency of power infrastructure and cooling devices by integrating them in the energy-aware resource management technique applied with the IT equipment.

## Author's Biography

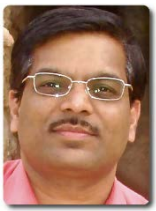

**Rajkumar Buyya** is a Redmond Barry Distinguished Professor and Director of the Cloud Computing and Distributed Systems (CLOUDS) Laboratory at the University of Melbourne, Australia. He is recognized as a "Web of Science Highly Cited Researcher" both in 2016 and 2017, a Fellow of IEEE, and Scopus Researcher of the Year 2017 with Excellence in Innovative Research Award for his outstanding contributions to Cloud computing. Contact him at rbuyya@unimelb.edu.au.

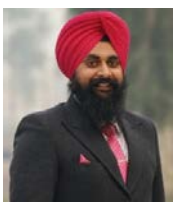

**Sukhpal Singh Gill** is a Postdoctoral Research Fellow with the University of Melbourne's Cloud Computing and Distributed Systems Laboratory. Contact him at sukhpal.gill@unimelb.edu.au.